\pdfoutput=1
\documentclass[a4paper,fleqn,usenatbib]{mnras}

\usepackage[T1]{fontenc}
\usepackage{ae,aecompl}

\usepackage{graphicx}	
\usepackage{amssymb}	
\usepackage{wasysym}
\usepackage{gensymb}

\title[H$_2$S in the L1157-B1 Bow Shock]{H$_2$S in the L1157-B1 Bow Shock\thanks{Herschel is an ESA space observatory with science instruments provided by European-led Principal Investigator consortia and with important participation from NASA.}
}

\author[J.
 Holdship et al.]{Jonathan Holdship$^{1}$\thanks{E-mail:jrh@star.ucl.ac.uk}, Serena Viti$^{1}$, Izaskun Jimenez-Serra$^{1}$, Bertrand Lefloch$^{2,3}$,
\newauthor Claudio Codella$^{4}$, Linda Podio$^{4}$, Milena Benedettini$^{5}$, Francesco Fontani$^{4}$,
\newauthor Rafael Bachiller$^{6}$, Mario Tafalla$^{6}$, Cecilia Ceccarelli$^{2,3}$
\\
$^{1}$Department of Physics and Astronomy, University College London, Gower Street, London, WC1E 6BT\\
$^{2}$Univ. Grenoble Alpes, Institut de Plan\'etologie et d'Astrophysique de Grenoble (IPAG), 38401 Grenoble, France\\
$^{3}$CNRS, Institut de Plan\'etologie et d'Astrophysique de Grenoble (IPAG), 38401 Grenoble, France\\
$^{4}$INAF, Osservatorio Astrofisico di Arcetri, Largo E. Fermi 5, 50125 Firenze, Italy\\
$^{5}$INAF, Istituto di Astrofisica e Planetologia Spaziali, via Fosso del Cavaliere 100, I-00133 Roma, Italy\\
$^{6}$IGN, Observatorio Astron\'omico Nacional, Calle Alfonso XII, E-28014 Madrid, Spain}
\date{Accepted XXX. Received YYY; in original form ZZZ}

\pubyear{2016}

\begin{document}
\label{firstpage}
\pagerange{\pageref{firstpage}--\pageref{lastpage}}

\maketitle

\begin{abstract}
Sulfur-bearing molecules are highly reactive in the gas phase of the ISM. However, the form in which most of the sulfur is locked onto interstellar dust grains is unknown. By taking advantage of the short time-scales of shocks in young molecular outflows, one could track back the main form of sulfur in the ices. In this paper, six transitions of H$_2$S and its isotopologues in the L1157-B1 bowshock have been detected using data from the Herschel-CHESS survey and the IRAM-30m ASAI large program. These detections are used to calculate the properties of H$_2$S gas in L1157-B1 through use of a rotation diagram and to explore the possible carriers of sulfur on the grains. The isotopologue detections allow the first calculation of the H$_2$S deuteration fraction in an outflow from a low mass protostar. The fractional abundance of H$_2$S in the region is found to be 6.0$\times$10$^{-7}$ and the deuteration fraction is 2$\times$10$^{-2}$. In order to investigate the form of sulfur on the grains, a chemical model is run with four different networks, each with different branching ratios for the freeze out of sulfur bearing species into molecules such as OCS and H$_2$S. It is found that the model best fits the data when at least half of each sulfur bearing species hydrogenates when freezing. We therefore conclude that a significant fraction of sulfur in L1157-B1 is likely to be locked in H$_2$S on the grains. 
\end{abstract}

\begin{keywords}
Stars: formation - radio lines: ISM - submillimetre:ISM - ISM: molecules
\end{keywords}



\section{Introduction}
Low mass protostars drive fast jets during the earliest stages of their evolution. The collision between these jets and the protostar's parent cloud is supersonic, producing shock fronts of warm, dense gas. This in turn drives processes that greatly increase the chemical complexity, such as endothermic reactions, sputtering from dust grains and ice mantle sublimation. Many molecular species are enhanced in abundance through this process \citep{draine1983,vandishoeck1998,codella2013} and the observation and modelling of these species is a powerful tool for understanding their gas-phase chemistry and the initial preshock composition of the ices on the grains.\par
L1157-mm is a Class 0, low mass protostar driving a molecular outflow \citep{gueth1997} and is at a distance of 250 pc \citep{looney2007}. The outflow was discovered by \citet{umemoto1992} who also inferred the presence of strong shocks from the temperature of the gas implied by NH$_3$ and SiO emission. Since then, several bow shocks along the outflow have been identified and studied due to their rich chemical content. The brightest of these, L1157-B1, has a complex, clumpy structure \citep{tafalla1995,benedettini2007} and an age of order 1000 yr constrained by observations \citep{gueth1998} and models \citep{flower2012}. Emission from molecules thought to be formed on icy mantles such as methanol and water have been observed \citep{bachiller1997,busquet2014}. This high degree of study of the L1157 outflow has led to it being considered the prototype of a chemically rich molecular outflow \citep{bachiller2001}. Therefore, L1157 represents the best laboratory to study the initial composition of the ices.\par
Through the IRAM large program ASAI (Astrochemical Surveys At IRAM) and the Herschel Key project CHESS (Chemical HErschel Surveys of Star forming regions), a spectral survey of L1157-B1 has been performed in the frequency ranges of 80-350 GHz (IRAM-30m/EMIR) and 500-2000 GHz (Herschel-HIFI). The first results of the CHESS survey \citep{codella2010,lefloch2010} confirmed the chemical richness of the bow shock. Constraints on the gas properties were found using CO. \citet{lefloch2012} found the line profiles could be fitted by three velocity components of the form $I(v) \propto exp(-|v/v_0|)$, tracing three kinematic components of the gas (see Sect.~\ref{sec:origin}). \citet{gomez2015} found two of these components could fit the CS emission from L1157-B1. An LVG analysis of water emission detected by Herschel-HIFI indicated one of these components could explain most of the water emission though the higher E$_u$ transitions required a hot component that had not been identified in previous studies \citep{busquet2014}. This implies that these velocity components likely represent distinct physical conditions in the gas. Several studies have started to compare the observed species with chemical models to derive the properties of the shock in L1157-B1 \citep{viti2011,benedettini2012,benedettini2013}. \citet{podio2014} reported the emission of molecular ions, including SO$^+$ and HCS$^+$, whose chemistry is closely related to the amount of S-bearing molecules released from dust grains, such as H$_2$S and OCS.\par
Sulfur is highly reactive and therefore its chemistry is very sensitive to the thermal and kinetic properties of the gas. That chemistry, if it were well understood, could be used to constrain the properties of the bow shock or regions within it. In fact, H$_2$S has been studied along with SO and SO$_2$ as a chemical clock for environments where dust grains are disrupted since it was first proposed by \citet{charnley1997}. A key assumption for this is that H$_2$S is one of the most abundant sulfur bearing species on ice mantles \citep{wakelam2004} and is then released into the gas phase during a shock. This assumption requires further study to determine its veracity. OCS and SO$_2$ are the only sulfur-bearing species firmly detected in the solid phase so far \citep{geballe1985,palumbo1995,boogert1997}. In fact, the chemical model presented in \citet{podio2014} in order to reproduce the abundance of CS and HCS$^+$ observed in L1157-B1 indicated that OCS could be one of the main sulfur carriers on dust grains. Despite this, detections of solid OCS give abundances that would account for only $\sim$0.5\% of the cosmic sulfur abundance \citep{palumbo1997}. Since estimates of grain-surface OCS abundances may be affected by blending with a nearby methanol feature in infra-red spectra, it is possible that much sulfur is eventually locked in solid OCS.\par
Other forms of sulfur have been considered. \citet{woods2015} investigated the gas and solid-phase network of sulfur-bearing species using new experimental and theoretical studies \citep{garozzo2010, ward2012, loison2012}. They found that although species such as H$_2$S and H$_2$S$_2$ are present on the ices, they are not sufficiently abundant to be the major carriers of sulfur. In fact, a residue of atomic sulfur may be abundant enough to harbor a large amount of the total sulfur budget. This residue may come off during sputtering if a fast enough shock is present. A shock velocity of 25 km s$^{-1}$ is likely to be sufficient to release even highly refactory species from the grains \citep{may2000}. One would expect to commonly observe chemistry involving atomic sulfur in shocked regions as this sulfur residue is sputtered from the mantle at such shock velocities. They also found that H$_2$S is in fact the predominant sulfur-bearing species at the early phases of the pre-stellar core.\par
In summary, H$_2$S is routinely observed in warm, sometimes shocked environments, in the gas phase; yet it is undetected in the ices to date. Given the uncertainty of the origin of H$_2$S on grains and the fact that L1157-B1 contains gas that is highly processed by shocks, the physical properties and kinematics of the H$_2$S gas could help determine its chemical origin. \par
In this paper we present H$_2$S observations along the L1157-B1 clump and by the use of a gas-grain chemical and parametric shock model we attempt to discover its origin and its relationship to other sulfur-bearing species. In Sect. 2, the observations with HIFI and the IRAM 30m telescope are described. In Sect. 3, the detected H$_2$S lines are presented and analyzed. In Sect. 4, the abundance and behaviour of H$_2$S is compared to CS and NH$_3$ as well as to a chemical model in which the form of sulfur in ice mantles is explored. Finally, the paper is summarized in Sect. 5.

\section{Observations}
\label{sec:observations}
\subsection{IRAM-30m}
\begin{figure}
\includegraphics[width=0.5\textwidth]{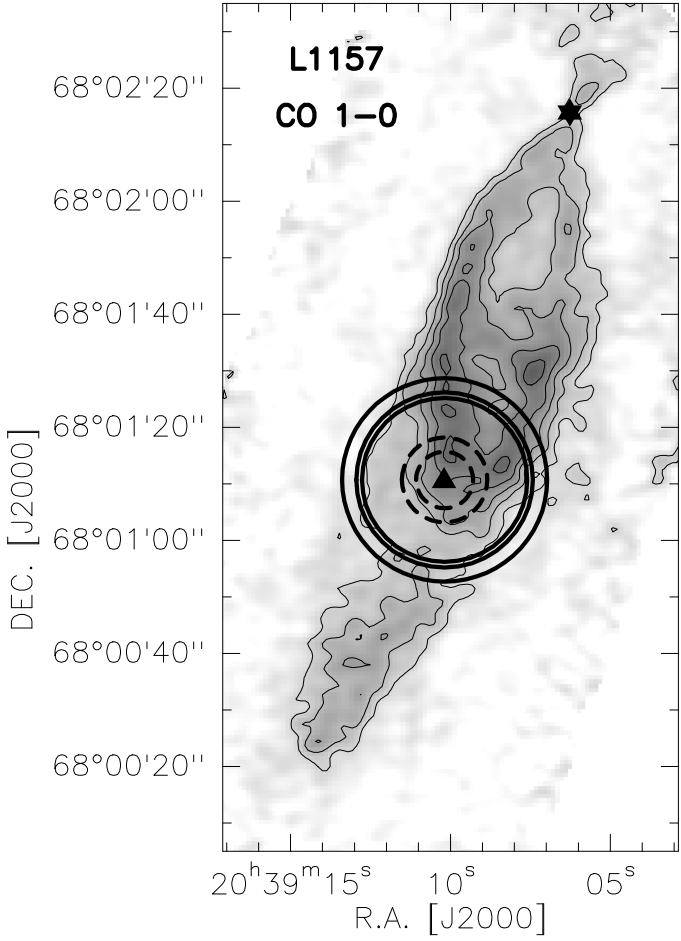}
\caption{CO (1-0) map of the L1157 region \citep{gueth1996}, including the protostar L1157-mm and the southern outflow. The dashed and solid rings show the beamsizes of IRAM-30m/EMIR and Herschel/HIFI respectively for the transitions detected. The triangle within the rings marks the position of the B1 shock and the other triangle marks the position of the protostar L1157-mm.}
\label{fig:map}
\end{figure}
The H$_2$S and H$_2^{34}$S (1$_{1,0}$-1$_{0,1}$) lines as well as the HDS (1$_{1,0}$-0$_{0,0}$) were detected during the ASAI program's unbiased spectral survey of L1157-B1 with the IRAM 30m telescope. The pointed co-ordinates were $\alpha _{J2000} =20^h39^m10.2^s$, $\delta _{J2000} =+68^{\degree}01'10".5$ i.e. $\Delta\alpha=+25".6 $, $\Delta\delta=-63".5$ from the driving protostar. NGC7538  was used for pointing, which was monitored regularly. It was found very stable with corrections less than 3". The observed region and the relevant beamsizes for the observations presented here are shown in Fig.~\ref{fig:map}. The broadband EMIR receivers were used with the Fourier Transform Spectrometer backend which provides a spectral resolution of 200 kHz, equivalent to 0.35 km s$^{-1}$ for the transitions at around 168 GHz. The transition at 244 GHz was smoothed to a velocity resolution of 1.00 km s$^{-1}$ to improve the signal to noise ratio. Forward and beam efficiencies are provided for certain frequencies \citep{kramer2013} and a linear fit gives a good estimate when interpolating between points. Table~\ref{table:telescope} gives these efficiencies as well as the telescope half power beamwidth at the observed frequencies. Line intensities were converted to units of T$_{MB}$ by using the B$_{eff}$ and F$_{eff}$ values given in Table~\ref{table:telescope}.

\subsection{Herschel-HIFI}
The H$_2$S (2$_{0,2}$-1$_{1,1}$), (2$_{1,2}$-1$_{0,1}$) and HDS (2$_{1,2}$-1$_{0,1}$) transitions were detected with Herschel \citep{pilbratt2010} using the HIFI instrument \citep{degraauw2010}, in bands 2A, 2B and 1B respectively as part of the CHESS spectral survey. The observations were done in double beam switching mode, with the pointed co-ordinates being the same as the IRAM-30m observations. The receiver was tuned in double sideband and the Wide Band Spectrometer was used. At the frequencies observed, the pointing of the H and V polarisations can differ by as much as 5". However, the spectra are in good agreement having similar rms values, line profiles and fluxes for detected lines and are therefore averaged. The data were processed using ESA supported package HIPE 6 (Herschel Interactive Processing environment, \citealt{ott2010}) and fits files from level 2 were exported for further reduction in GILDAS\footnote{http://www.iram.fr/IRAMFR/GILDAS/}. The resulting spectra have a velocity resolution of 0.25 km s$^{-1}$ and an rms of 3.6 mK. Values for beam efficiency were taken from \citet{roelfsema2012} and adjusted for wavelength by a Ruze formula given in the same article.
\begin{figure}
\includegraphics[width=0.5\textwidth]{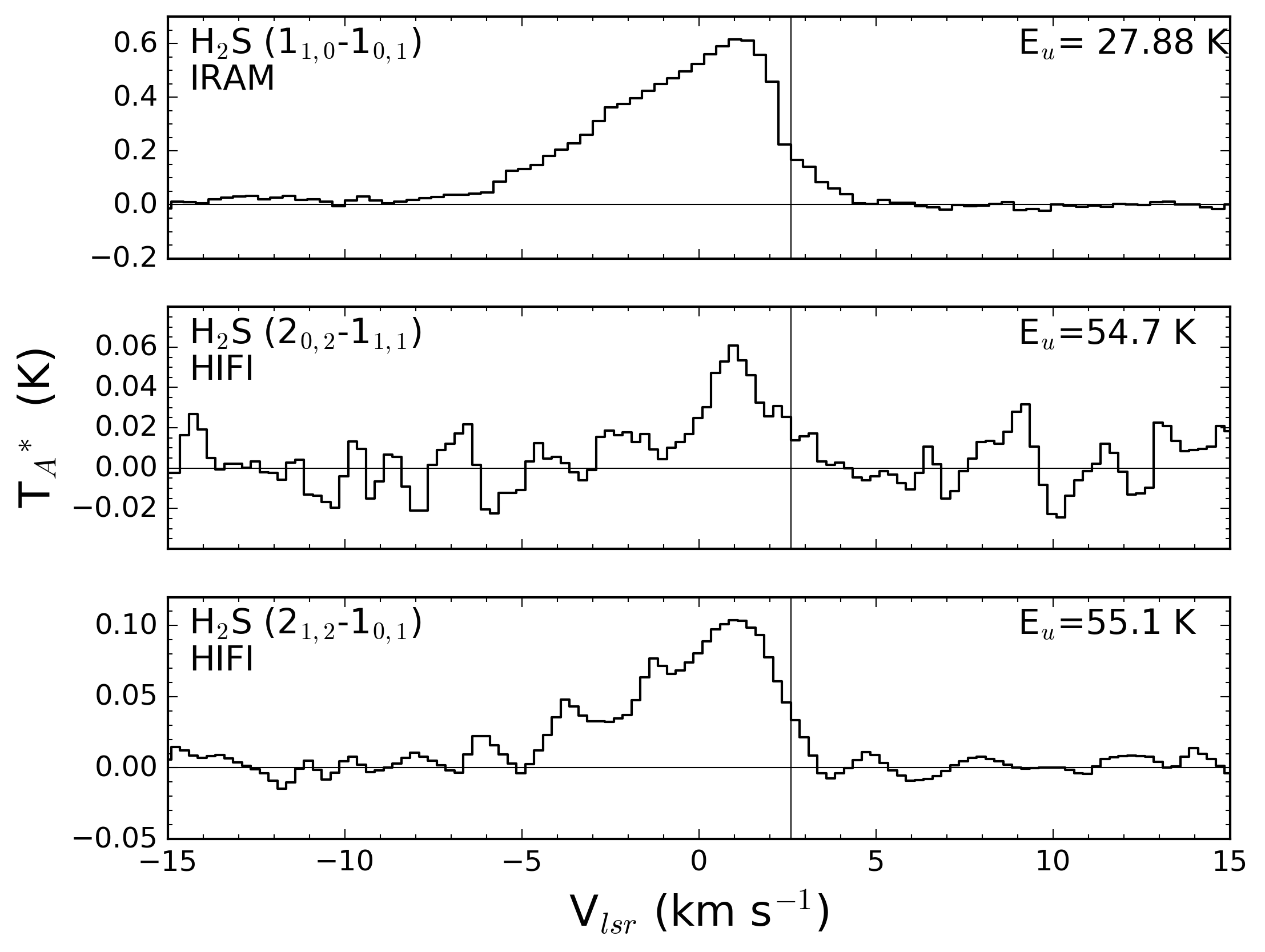}
\caption{Herschel/HIFI and IRAM-30m/EMIR detections of the H$_2$S main isotopologue. The vertical line indicates the cloud velocity V$_{sys}=$2.6 km s$^{-1}$.}
\label{fig:h2sspectra}
\end{figure}
\begin{figure}
\includegraphics[width=0.5\textwidth]{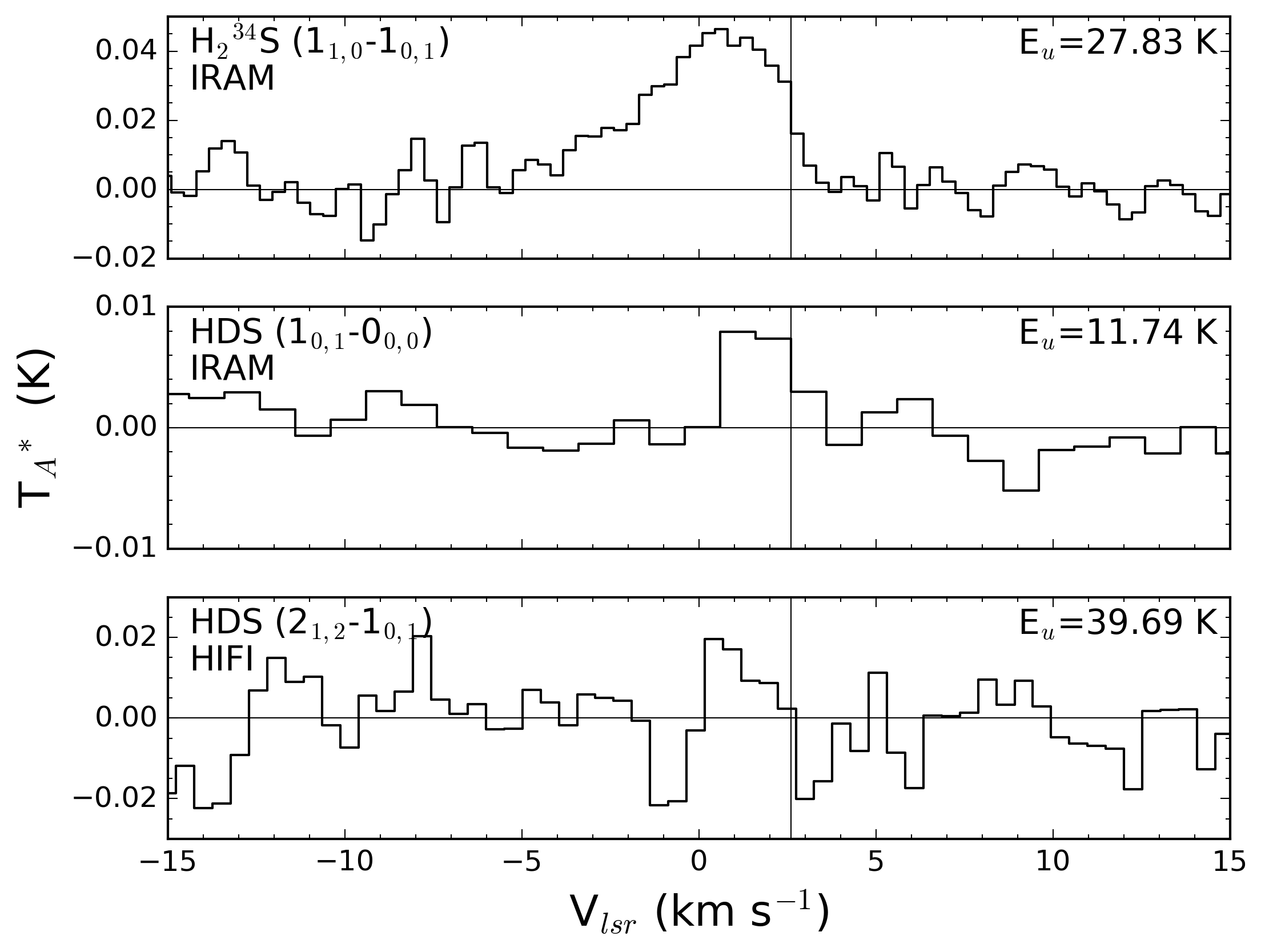}
\caption{Herschel/HIFI and IRAM-30m/EMIR detections of H$_2$S isotopologues. The HDS lines are at lower spectral resolution than the other detections. The HDS (1-0) spectrum has a resolution of 1 km s$^{-1}$ and HDS (2-1) spectrum has a resolution of 0.5 km s$^{-1}$.}
\label{fig:isospectra}
\end{figure}

\begin{table}
\caption{Details of Herschel/HIFI observations.}
\begin{tabular}{cccc}
\hline
\textbf{Frequency} & \textbf{Band} & \textbf{Obs\_Id} & \textbf{Date} \\
\textbf{(GHz)} & & & \\
\hline
687.3  & 2a & 1342207607 & 2010-10-28 \\
736.0  & 2b & 1342207611 & 2010-10-28 \\
582.4  & 1b & 1342181160 & 2009-08-01 \\
\hline
\end{tabular}
\label{table:hifi}
\end{table}

\begin{table*}
\centering
\caption{List of detected lines and the relevant spectroscopic and instrument properties. Ortho and para species are denoted by o- and p- respectively. All spectroscopic data were taken from \citet{pickett1998}.}
\begin{tabular}{cccccccccccc}
\hline
\textbf{Molecule} & \textbf{Transition} & \textbf{Freq} & \textbf{E$_u$} & \textbf{g$_u$} & \textbf{log(A$_{ul}$)} & \textbf{Instrument} &\textbf{T$_{sys}$} & \textbf{$\theta_{beam} $} & \textbf{$\Delta$V} & \textbf{B$_{eff}$} & \textbf{F$_{eff}$}    \\
&& \textbf{(GHz)} & \textbf{(K)} &&&&\textbf{(K)}&\textbf{(")} & \textbf{(km s$^{-1}$)}&& \\
\hline
o-H$_2$S      & 1$_{1,0}$-1$_{0,1}$ & 168.7628    & 27.9 & 9 & -4.57 & IRAM-30m/EMIR & 132 & 15 & 0.35 & 0.74 & 0.93  \\
p-H$_2$S      & 2$_{0,2}$-1$_{1,1}$ & 687.3035    & 54.7 & 5 & -3.03 & Herschel/HIFI & 137 & 31 & 0.25 &0.76 & 0.96  \\
o-H$_2$S      & 2$_{1,2}$-1$_{0,1}$ & 736.0341    & 55.1 & 15 & -2.88 & Herschel/HIFI & 317 & 29 & 0.25 &0.77 & 0.96  \\
o-H$_2^{34}$S & 1$_{1,0}$-1$_{0,1}$ & 167.9105     & 27.8 & 9 & -4.58 & IRAM-30m/EMIR & 159 & 15 & 0.36 &0.74 & 0.93  \\
o-HDS         & 1$_{0,1}$-0$_{0,0}$ & 244.5556    & 11.7 & 3 & -4.90 & IRAM-30m/EMIR & 190 & 10 & 1.00 &0.59 & 0.92  \\
o-HDS        & 2$_{1,2}$-1$_{0,1}$ & 582.3664    & 39.7 & 5 & -3.18 & Herschel/HIFI & 84& 36 & 0.50 &0.77 & 0.96  \\
\hline
\end{tabular}
\label{table:telescope}
\end{table*}
\begin{table*}
\centering
\caption{Derived parameters of the detected lines. T$_{peak}$ is given in units of T$_{mb}$ and Vmin/Vmax are where lines drop below 3$\sigma$ levels except for HDS lines where 1$\sigma$ is used (See Sect.~\ref{sec:optdepth}). Parenthesized values are uncertainties.}
\begin{tabular}{ccccccc}
\hline
\textbf{Molecule} & \textbf{Transition} & \textbf{T$_{peak}$} & \textbf{V$_{peak}$} &\textbf{V$_{min}$/V$_{max}$} & \textbf{$\int$ T$_{mb}$ dv} \\
&& \textbf{(mK)} & \textbf{(km s$^{-1}$)} &\textbf{(km s$^{-1}$)} & \textbf{(K km s$^{-1}$)} \\
\hline
o-H$_2$S      & 1$_{1,0}$-1$_{0,1}$ & 614 (8)& 1.05 & -7.90/4.70  &4.41 (0.90)\\
p-H$_2$S      & 2$_{0,2}$-1$_{1,1}$ & 60 (9) & 1.00 & 0.10/2.10   &0.20 (0.03)\\
o-H$_2$S      & 2$_{1,2}$-1$_{0,1}$ & 103 (9)& 1.25 & -4.15/3.10  &0.58 (0.07)\\
o-H$_2^{34}$S & 1$_{1,0}$-1$_{0,1}$ & 46 (6) & 0.65 & -1.69/2.96  &0.28 (0.06)\\
o-HDS         & 1$_{0,1}$-0$_{0,0}$ & 8 (2)  & 1.00 & 0.60/3.60   &0.030 (0.008)\\
o-HDS         & 2$_{1,2}$-1$_{0,1}$ & 19 (6) & 0.00 & 0.00/2.50   &0.034 (0.009)\\
\hline
\end{tabular}
\label{table:measured}
\end{table*}
\section{Results}
\label{sec:results}
\subsection{Detected Lines and Opacities}
\label{sec:optdepth}
In total, six lines were detected including H$_2^{34}$S and HDS as well as the main isotopologue, H$_2$S. Table~\ref{table:telescope} gives the telescope and spectroscopic parameters relevant to each detected line. Frequencies ($\nu$), quantum numbers, Einstein coefficients (A$_{ul}$), and the upper state energies (E$_u$) and degeneracies (g$_u$) are taken from the JPL spectral line catalog \citep{pickett1998}. The measured line properties including the velocities V$_{min}$ and V$_{max}$ taken to be the velocities at which the emission falls below the 3$\sigma$ level, and the integrated emission ($\int T_{MB}dv$) within those limits are shown in Table~\ref{table:measured}. The error quoted for the integrated intensities includes the propagated error from the rms of T$_{MB}$ and the velocity resolution of the spectra as well as nominal calibration errors of 10 and 20\% for Herschel-HIFI and IRAM-30m respectively. Note that the HDS spectra are at lower spectral resolution and have less significant detections than the other H$_2$S transitions, each peaking between the 3 and 4 $\sigma$ level. For this reason, the velocity limits of the HDS (2-1) transition are taken to be where the peak is above 1$\sigma$.\par
Figures~\ref{fig:h2sspectra} and \ref{fig:isospectra} show the six lines labelled by their species and quantum numbers. The spectra are given in units of antenna temperature T$_a^*$. Figure~\ref{fig:h2sspectra} shows the three H$_2$S lines whilst Fig.~\ref{fig:isospectra} shows the isotopologues. In Fig.~\ref{fig:h2sspectra}, the H$_2$S (2$_{1,2}$-1$_{0,1}$) line shows three peaks. The primary peak is at 1.25 km s$^{-1}$, which is common to all the spectra. The secondary peak at -3.75 km s$^{-1}$ is consistent with lines detected in L1157-B1 by \citet{codella2010} who found that HCN, NH$_3$, H$_2$CO and CH$_3$OH showed primary peaks at approximately 1 km s$^{-1}$ and secondary peaks between -3 and -4 km s$^{-1}$. However, the peak at -1.25 km s$^{-1}$ in the H$_2$S (2$_{1,2}$-1$_{0,1}$) spectrum is not consistent with any other spectral features and may be due to a contaminant species or simply noise.\par
The 1$_{1,0}$-1$_{0,1}$ transition has been detected in both H$_2$S and H$_2^{34}$S, allowing the optical depth of L1157-B1 for H$_2$S to be calculated. Equation~\ref{Nu} gives the source averaged column density of the upper state of a transition, $N_u$, when the source is much smaller than the beam and optical depth, $\tau$ is non-negligible. $ff$ is a correction for the fact the emission does not fill the beam.
\begin{equation}
\label{Nu}
N_u = \frac{8\pi k\nu^2}{hc^3A_{ul}}ff\int T_{MB}dv\left(\frac{\tau}{1-e^{-\tau}}\right)
\end{equation}
It is assumed that all the H$_2$S emission comes from B1. Whilst previous work on L1157-B1 has suggested most emission comes from the walls of the B1 cavity \citep{benedettini2013}, this has not been assumed and the entire size of B1 is used. Thus, size was taken to be 18" in agreement with the size of B1 in CS estimated by \citet{gomez2015} from PdBI maps by \citet{benedettini2013}. The validity of this assumption is explored in Sect.~\ref{sec:origin}. However, it should be noted that varying the size from 15" to 25" changes ln($\frac{N_u}{g_u}$) by less than 1\% and encompasses all estimates of the source size from other authors \citep{lefloch2012,podio2014,gomez2015}. \par
For the equivalent transition in a pair of isotopologues, it is expected that $N_u$ would only differ by the isotope abundance ratio if the emission region is the same. A $^{32}$S/$^{34}$S ratio, R, of 22.13 was assumed \citep{rosman1998} and tested by comparing the integrated emission in the wings of the 1$_{1,0}$-1$_{0,1}$ transitions; that is all of the emission below -1 km s$^{-1}$. This emission is likely to be optically thin, allowing us to test the abundance ratio. A ratio of 22$\pm$3 was found, consistent with the value taken from the literature. Using Eq.~\ref{intensity} below with the measured integrated line intensities, an optical depth for H$_2$S (1$_{1,0}$-1$_{0,1}$) of $\tau =$0.87 was calculated. It is assumed that H$_2^{34}$S is optically thin.
\begin{equation}
\label{intensity}
\frac{1-e^{-\tau}}{1-e^{-\tau/R}} = \frac{\int T_{MB}(^{32}S)dv}{\int T_{MB}(^{34}S)dv}
\end{equation}

\begin{figure*}
\centering
\includegraphics[width=\textwidth]{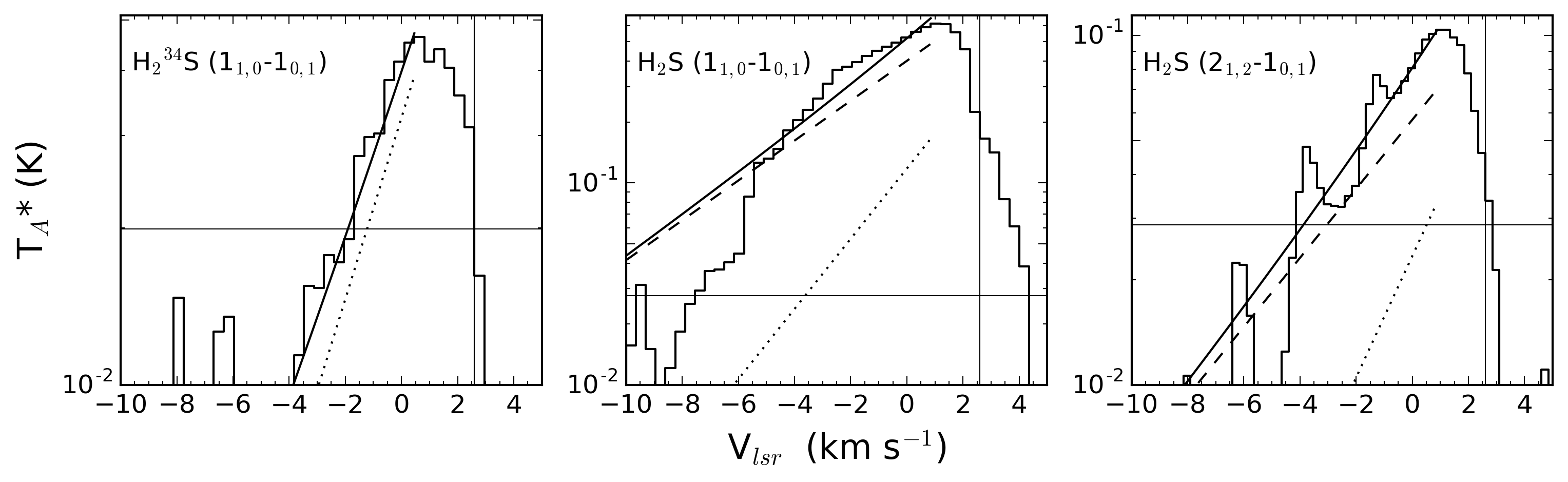}
\caption{Log-linear plots of the well detected transitions. The dashed and dotted fits are $I(v) \propto exp(-|v/v_0|)$ profiles with v$_0$ = 4.4 km s$^{-1}$ (g$_2$) and 2.5 km s$^{-1}$ (g$_3$) respectively. The solid line is the best fit line, a linear combination of the other two. The third component, g$_1$ (v$_0$=12.5 km s$^{-1}$) is not shown as it is not required for the best fit and there is no emission from velocities more blue shifted than -8 km s$^{-1}$. The vertical line is the cloud velocity, V$_{sys}=$2.6 km s$^{-1}$ \citep{bachiller1997}, and the horizontal line marks each the 3$\sigma$ rms level on each plot.}
\label{fig:components}
\end{figure*}
\subsection{Origin of the Emission}
\label{sec:origin}
\citet{lefloch2012} found that the CO emission from L1157-B1 could be fitted by a linear combination of three velocity profiles, associated with three physical components. The profiles were given by $I(v) \propto exp(-|v/v_0|)$ where $v_0$ is the characteristic velocity of the physical component. The velocities were 12.5 km s$^{-1}$, 4.4 km s$^{-1}$ and 2.5 km s$^{-1}$ which are respectively associated with a J shock where the protostellar jet impacts the B1 cavity, the walls of the B1 cavity and an older cavity (B2). These components were labelled as g$_1$, g$_2$ and g$_3$ and the same notation is used here for consistency. This has also been applied to other molecules. For CS \citep{gomez2015}, it was found that the g$_2$ and g$_3$ components fit the CS lines well with a negligible contribution from the g$_1$ component except for the high J lines detected with HIFI. The same was found to be true for SO$^+$ and HCS$^+$ \citep{podio2014}. This implies that the majority of emission from the low energy transitions of sulfur bearing molecules comes from the B1 and B2 cavities and is not associated with the J-shock.\par
The three components have been used to fit the H$_2$S line profiles and the results are shown in Fig.~\ref{fig:components}. The fit for H$_2$S (2$_{0,2}$-1$_{1,1}$) is not shown as the detection is not significant enough to be well fitted. Unlike other molecules in the region, the fits for H$_2$S appear to be fairly poor, with reduced $X^2$ values much greater than 1 for H$_2$S (1$_{1,0}$-1$_{0,1}$) and the H$_2$S (2$_{1,2}$-1$_{0,1}$) suffering from the secondary peaks. We also note that the H$_2$$^{34}$S isotopologue is well fitted. In each case, only the g$_2$ and g$_3$ components are required for the best fit. This implies that, as for CS, the majority of the H$_2$S emission arises from B1 and B2 cavities affected by the interaction of low-velocity C-shock. Since the g$_1$ component is negligible in the H$_2$S lines, the nearby J-shock does not contribute to their emission. This justifies the use of a C-shock model in Sect.~\ref{sec:model} to reproduce the observed line profiles and measured abundances of H$_2$S in L1157-B1. \par

\subsection{Column Densities and Abundances}
\label{sec:abund}
The three detected transitions of H$_2$S allow the excitation temperature and column density of the shocked gas in L1157-B1 to be calculated through use of a rotation diagram \citep{goldsmith1999}. Of the three transitions, one is para-H$_2$S and the other two are ortho. In order to proceed, an ortho-to-para ratio needs to be assumed. Given the lack of further data, the statistical average of 3 is assumed for the ortho-to-para ratio. Reducing this ratio to 1 produced results within error of those presented here for column density and temperature. It is therefore not possible to draw any conclusions about the ortho-to-para ratio from the data available. \par
\begin{figure}
\includegraphics[width=0.5\textwidth]{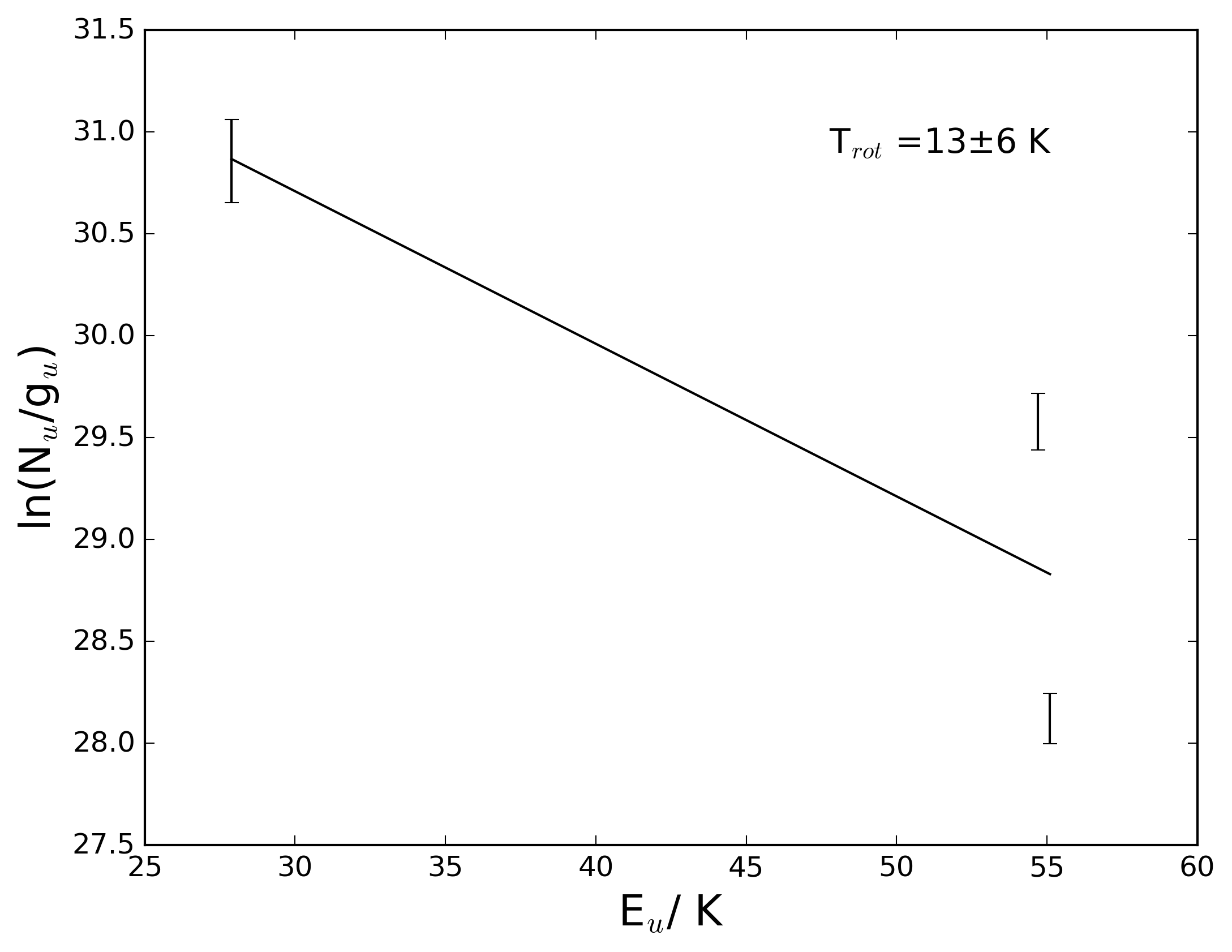}
\caption{Rotation diagram for H$_2$S. The gradient gives an excitation temperature of 13 K and the intercept gives a column density of 6.0$\pm$4.0 $\times$10$^{14}$ cm$^{-2}$ at that temperature.}
\label{fig:rotdiag}
\end{figure}

The H$_2$S column density and temperature were calculated through the use of a rotation diagram. For the rotation diagram, the upper state column density, $N_u$ was calculated from Eq.~\ref{Nu} assuming all three transitions had the same optical depth, calculated above. The filling factor in Eq.~\ref{Nu}, $ff$ is uncertain due to the fact that Sect.~\ref{sec:origin} shows that H$_2$S emission comes partially from the walls of the B1 cavity and partially from the older B2 cavity. For the calculations here, we adopt a source size of 18" but as noted in Sect.~\ref{sec:optdepth}, a size of 25" gave the same result. The rotation diagram is shown in Fig.~\ref{fig:rotdiag} and gave an excitation temperature of 13$\pm$6 K. This is consistent with the low temperature of the g$_3$ component calculated as $\sim$23 K by \citet{lefloch2012}. Using the JPL partition function values for H$_2$S \citep{pickett1998}, a total column density was found to be $N(H_2S)$=6.0$\pm$4.0$\times$10$^{14}$ cm$^{-2}$. \par
\citet{bachiller1997} calculated a fractional abundance of H$_2$S by comparing measured column densities of H$_2$S to CO and assuming a CO:H$_2$ ratio of 10$^{-4}$. They obtained a value of 2.8$\times$10$^{-7}$. Given that their H$_2$S column density was calculated from a single line by assuming a temperature of 80 K and optically thin emission, an updated value has been calculated. For the CO column density, a more recent measurement of N(CO)=1.0$\times$10$^{17}$ cm$^{-2}$ is used \citep{lefloch2012}. For a H$_2$S column density of 6.0$\times$10$^{14}$ cm$^{-2}$, a fractional abundance X= 6.0$\times$10$^{-7}$ is found. This is a factor of $\sim$2 larger than the \citeauthor{bachiller1997} measurement, most likely due to the improved column density derived by using a rotation diagram with three transitions rather than a single line with an assumed temperature.\par
It should be noted that the critical densities for the two higher frequency transitions are n$_c\sim 10^7cm^{-3}$. Estimates of the average number density in B1 and B2 are of order n$_H\sim10^5cm^{-3}$. Therefore LTE is unlikely to apply here. This casts doubt on the rotation diagram method and so the radiative transfer code RADEX \citep{vandertak2007} was run with a range of column densities, gas densities and temperatures to see if similar results would be obtained. H$_2$ number densities in the range $n = 10^{4} - 10^{6} cm^{-3}$ and temperatures in the range 10-70 K were used based on the values found for g$_2$ and g$_3$ with LVG modelling by \citet{lefloch2012}. With these parameters, the best fit found by comparing predicted brightness temperatures to the H$_2$S lines reported here is 3.0$\times10^{14}cm^{-2}$ at a temperature of 10 K and density 10$^{5}$cm$^{-3}$. Fits with T \textgreater~30 K or n$_H$ $\gg$ 10$^5$cm$^{-3}$ were always poor.
\subsection{The Deuteration Fraction of H$_2$S}
The excitation temperature is further used to make the first estimate of the deuteration fraction of H$_2$S in an outflow from a low mass protostar. The integrated intensities of the HDS emission lines are used to calculated $ln(N_u)$ for each transition, again assuming a source of 18". We then use T$_{ex}$=13$\pm$6 K, the value of the partition function, Z, and upper state degeneracy g$_u$ taken from the JPL catalog to find a column density for HDS as 
\begin{equation}
\label{roteq}
 ln(N)=ln(\frac{N_u}{g_u}) + ln(Z) + \frac{E_u}{kT_{ex}}
\end{equation} 
By averaging the column density obtained from each of the two transitions, a column density of 2.4$\pm$1.2$\times$10$^{13}$ cm$^{-2}$ is found. This corresponds to a deuteration fraction for H$_2$S of 2.5$\pm$2.5$\times$10$^{-2}$. As these abundances are an average over the beam, this is likely to be a lower limit of the true deuteration on the grains. However, it is comparable to level of deuteration measured for other species in L1157-B1. For example, the methanol deuteration fraction was found to be 2$\times$10$^{-2}$ by \citet{codella2012}. Further, \citet{fontani2014} detected HDCO, CH$_2$DOH with the Plateau de Bure interferometer and compared to previous measurements of CH$_3$OH and H$_2$CO. They obtain higher deuteration fractions for H$_2$CO and CH$_3$OH than the H$_2$S value reported here. However, they note that the PdBI detections of the H$_2$CO and CH$_3$OH that they use may be resolving out as much as 60\% of the emission, leading to higher deuteration fractions and so claim consistency with the \citet{codella2012} results. Therefore, deuterated H$_2$S appears to be consistent with many deuterated molecules in L1157-B1. 

\section{Sulfur Chemistry in the B1 Shock}
\subsection{Comparison of Line Profiles}
\begin{figure}
\centering
\includegraphics[width=0.5\textwidth]{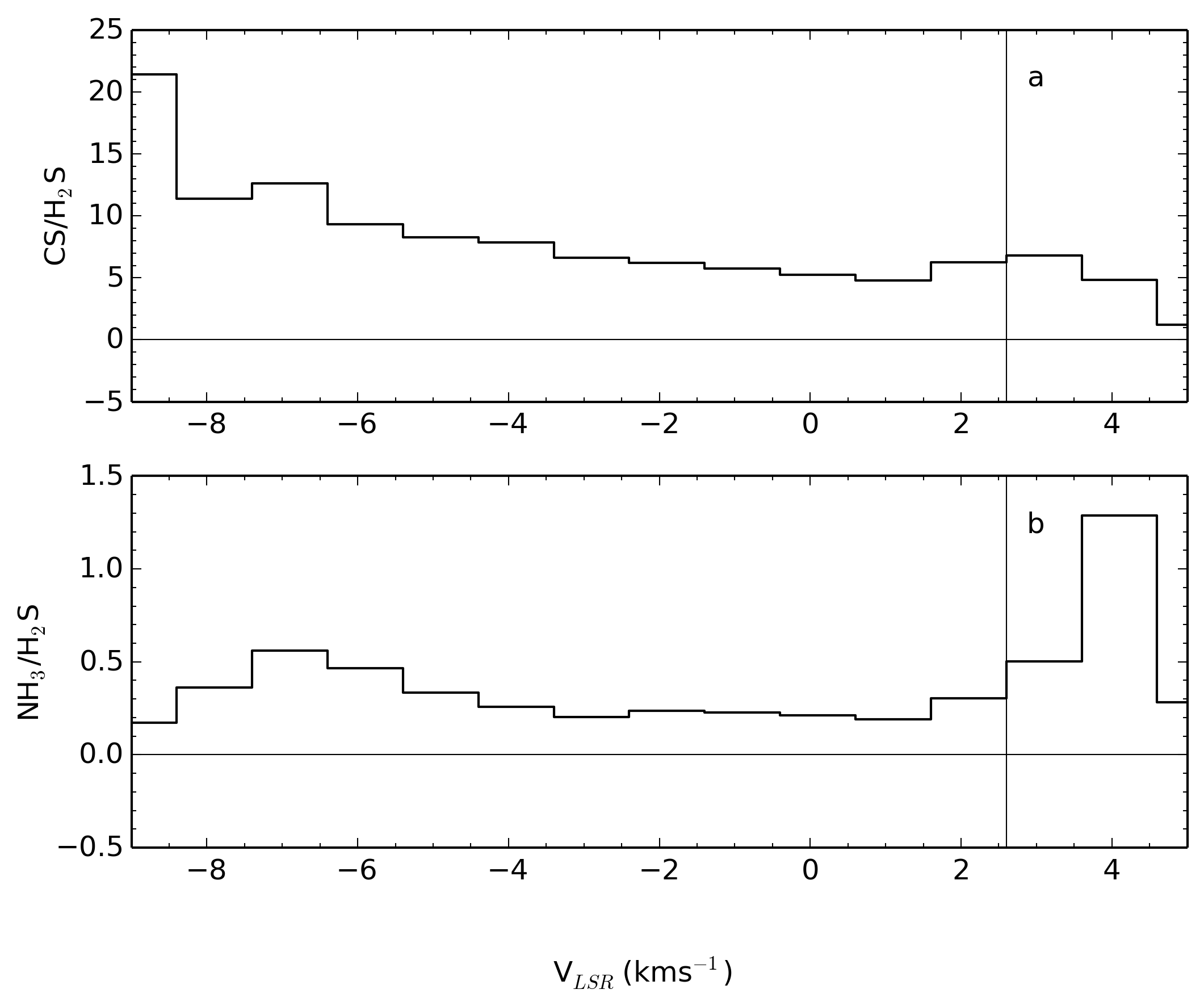}
\caption{Emisson ratios CS (3-2)/H$_2$S (a) and NH$_3$ (1-0)/H$_2$S (b). The H$_2$S (1$_{1,0}$-1$_{0,1}$) line is used in both. All the spectra have been resampled into bins with a common resolution of 1 km s$^{-1}$ centred at 2.6km s$^{-1}$. The 3$\sigma$ velocity limits of the H$_2$S line are used to set the x-axis as they give the smallest velocity range in each case.} 
\label{fig:ratios}
\end{figure}

In the previous section we postulate that H$_2$S arises, like CS, from the B1 and B2 cavity. In order to test this hypothesis, we compar the emission of these species to see if they show similar behaviour. Figure~\ref{fig:ratios} shows the ratios of the line temperatures for the H$_2$S (1$_{1,0}$-1$_{0,1}$) transition with the CS (3-2) and NH$_3$ (1-0) transitions. This allows us to better compare the line profiles and the evolution of the molecular abundances of the different species considered in this section and in Sect.~\ref{sec:model}. The spectra have been re-sampled to common velocity bins. The H$_2$S (1$_{1,0}$-1$_{0,1}$) transition was chosen as it is both less noisy than the other H$_2$S transitions and closer in excitation properties to the available CS and NH$_3$ transitions.\par
The CS transition in Fig.~\ref{fig:ratios}a is the CS (3-2) transition at 146.97 GHz; it has an upper level energy of E$_u =$ 14.1 K. Figure~\ref{fig:ratios}a shows that the CS (3-2) and H$_2$S (1$_{1,0}$-1$_{0,1}$) profiles differ by a factor of $\sim$10 at higher velocities. Assuming that the line ratio follows the behaviour of the abundance ratio between these two species, this may imply that H$_2$S is not as abundant as CS at higher velocities and therefore raises the possibility that the H$_2$S and CS emission do not entirely arise from the same region. This pattern is attributed to chemical differences between the molecules and explored in Sect.~\ref{sec:model}.\par
The NH$_3$ spectrum in Fig.~\ref{fig:ratios}b is the NH$_3$ (1-0) line at 572.49 GHz, it has an upper level energy of E$_u =$ 27.47 K which is comparable to the upper level energy of the H$_2$S (1$_{1,0}$-1$_{0,1}$) transition. These two transitions are compared due to their similar excitation properties and strong detections; the other H$_2$S spectra are much noisier. From Fig.~\ref{fig:ratios}b, we find that the H$_2$S and NH$_3$ intensities (and thus, their abundances) differ by less than a factor of 3 throughout the postshock gas, which suggests that H$_2$S and NH$_3$ come from the same region and behave similarly. We however note that the NH$_3$ line has been measured within a larger beam than H$_2$S and thus, the NH$_3$ line may include emission from a larger area in the L1157-B1 bowshock.
\subsection{Comparison with Chemical Models}
\label{sec:model}
We have established that the H$_2$S spectra do not require a g$_1$ component to be well fitted and are therefore not likely to be associated with the J shock from L1157-mm's outflow (see Sect.~\ref{sec:origin}). Extensive efforts have been made to model the dynamics  of L1157-B1 with C-shocks \citep[e.g.][]{gusdorf2008,flower2012} and have proven successful. The line profiles of H$_2$S appear to be consistent with the shapes predicted for other molecules from a parameterized C-shock model in \citet{jimenez2009} and the behaviour of other species such as H$_2$O and NH$_3$ have been successfully modelled by coupling the same shock model with a chemical code \citep{viti2011}. Hereafter we thus assume that the sulfur chemistry observed toward L1157-B1 can be modelled by a C-type shock. Through this modelling, we investigate whether H$_2$S is formed mainly on the grains or in the gas phase during the passage of a C-type shock. We shall use the same model as in \citet{viti2011}, \textsc{ucl\_chem}, which is a gas-grain chemical model coupled with a parametric shock model \citep{jimenez2008}. The model is time-dependent and follows the abundances of molecular species through two phases. \par
In phase I, the formation of a molecular cloud is modelled including gravitational collapse, freeze out onto dust grains and surface chemistry. This gives a self consistent rather than assumed ice mantle and gas phase abundances for phase II. \par
In phase II, the propagation of a C-type shock from the protostellar outflow is modelled according to the \citet{jimenez2008} parameterization. The physical structure of the shock is parameterized as a function of the shock velocity (V$_s$), and the H$_2$ gas density of the preshock gas. The magnetic field used in the model is 450 $\mu$G, transverse to the shock velocity, and it has been calculated following the scaling relation B$_0$($\mu$G) = b$_0$ x $\sqrt{n_0 (cm^{-3})}$ with b$_0$ = 1 following \citet{draine1983} and \citet{bourlot2002}. Note that in this one case, the density n$_0$ is given as the density of hydrogen nuclei. The maximum temperature of the neutral fluid attained within the shock (T$_{n,max}$) is taken from Figures 8b and 9b in \citet{draine1983}. The sputtering of the grain mantles is simulated in the code by introducing a discontinuity in the gas-phase abundance of every molecular species once the dynamical time across the C-shock has reached the ``saturation time-scales''. The ``saturation times'' were defined by \citet{jimenez2008} as the time-scales at which the relative difference in the sputtered molecular abundances between two consecutive time steps in the model are less than 10\%, and represent a measure of the time-scales for which almost all molecular material within the mantles is released into the gas phase. For the saturation times for H$_2$S and the other sulfur-bearing species, we adopt those already derived for SiO coming from the mantles (see Table 5 in \citet{jimenez2008}). The mathematical expression for the sputtering of the grain mantles for any molecular species is the same except for the initial solid abundance in the ices (see Eqs. B4 and B5 in the same paper). Consequently, species differ in the absolute scale in the final abundance sputtered from grains but not the time-scales at which they are injected into the gas phase. \par
We have used the best fit model from \citet{viti2011}. In this model, the C-shock has a speed of v$_s$=40 km s$^{-1}$ and the neutral gas reaches a maximum temperature of T$_{n,max}$=4000 K in the postshock region. As shown by \citet{bourlot2002} and \citet{flower2003}, C-shocks can indeed develop in a medium with a pre-shock density of 10$^5$ cm$^{-3}$ and magnetic induction B$_0$=450 $\mu$G (or b$_0$=1; see their Figures 1 and 5). Although the terminal velocity of the H$_2$S shocked gas observed in L1157-B1 is only about -10 km s$^{-1}$, the higher shock velocity of vs= 40 km s$^{-1}$ used in the model is justified by the fact that other molecules such as H$_2$O and CO show broader emission even at terminal velocities of -30 km s$^-1$ and -40 km s$^{-1}$ \citep{codella2010,lefloch2010}. These are expected to be more reflective of the shock velocity as these molecules are not destroyed at the high temperatures produced by the shock and so remain abundant out to high velocities \citep{viti2011,gomezruiz2016}. \par
We note that recent shock modelling assuming perpendicular geometry of the shock predicts thinner postshock regions, and consequently higher T$_{max}$ in the shock, than those considered here under the same initial conditions of preshock gas densities and shock velocities \citep[see e.g.][]{guillet2011,anderl2013}. However, MHD simulations of oblique shocks \citep{vanloo2009} seem to agree with our lower estimates of T$_{n,max}$ and therefore, we adopt the treatment proposed by \citet{jimenez2008}. The validation of this parameterization is discussed at length in \citet{jimenez2008}. The interested reader is referred to that paper for details of the C-shock modelling and to \citet{viti2004} for details on \textsc{ucl\_chem}.\par
In the model, we have expanded the sulfur network following \citet{woods2015} to investigate the composition of the ice mantles. The first three versions of the network were taken directly from \citet{woods2015}. The behaviour of sulfur bearing species are as follows (see Table~\ref{tab:networks}):
\begin{itemize}
\item A \- Species froze as themselves.
\item B \- Species immediately hydrogenated.
\item C \- Species would freeze 50\% as themselves and 50\% hydrogenated. 
\item D \- Any species for which a reaction existed to produce OCS on the grains was set to freeze out entirely as OCS.
\item E \- Species would freeze 50\% as themselves and 50\% as OCS if possible.
\end{itemize}
The injection of the ice mantles into the gas phase by sputtering occurs once the dynamical time in the C-shock exceeds the saturation time, i.e. when t$_{dyn}$ \textgreater 4.6 years. No grain-grain interactions are taken into account due to the computational costs.\par
The parameter values for the C-shock in the model are given in Table~\ref{table:models}. Note network D is not reflective of real chemistry but rather takes the extreme case of highly efficient grain surface reactions allowing frozen sulfur, carbon and oxygen atoms to form OCS.\par
\begin{figure*}
\centering
\includegraphics[width=\textwidth]{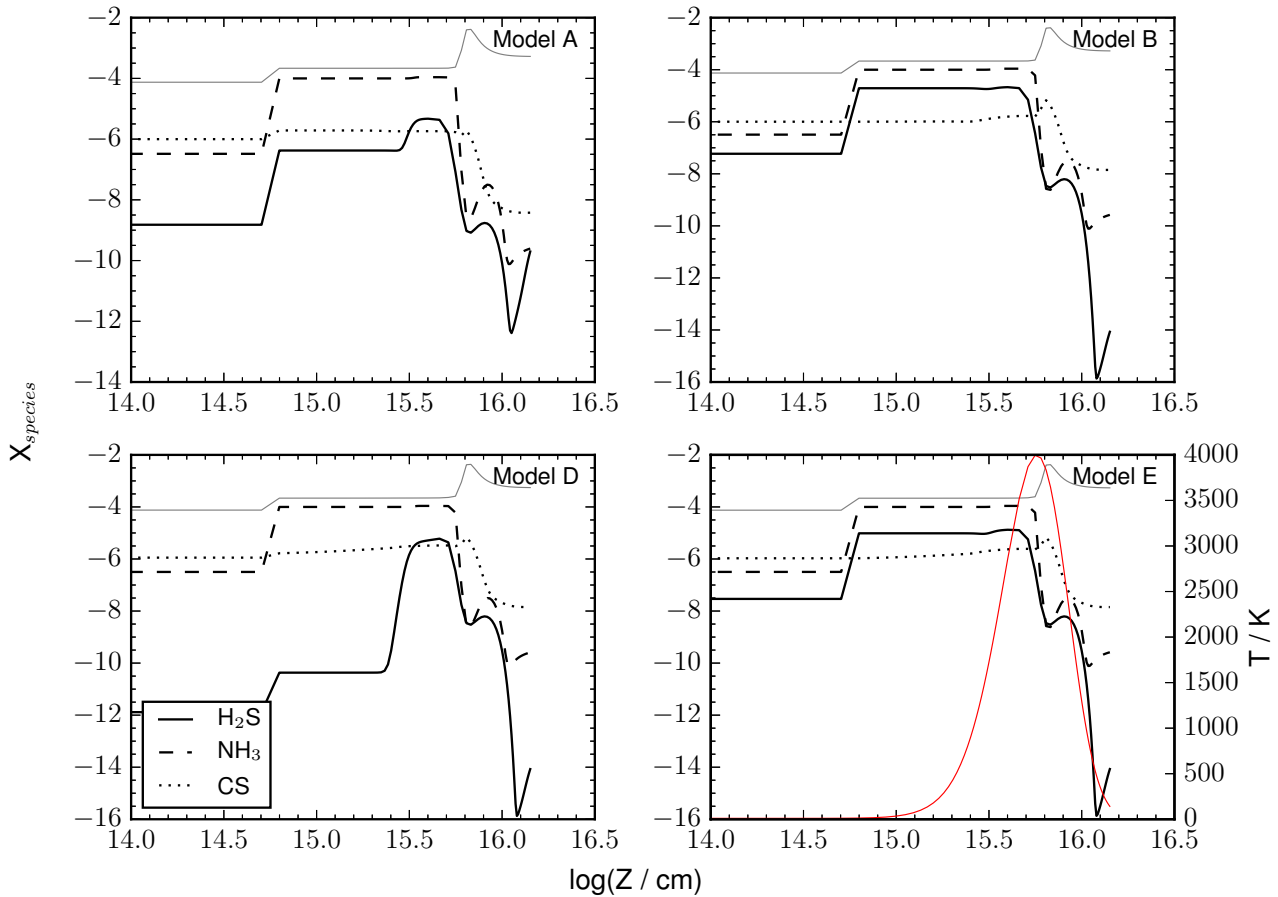}
\caption{Fractional abundances of H$_2$S, NH$_3$ and CS as a function of distance into the shock for the different networks. Plots are shown only up to the dissipation length, described in Section~\ref{sec:model}. The panels show: A) Freeze out to same species; B) freeze out to hydrogenated species; D) efficient OCS freeze out; E) 50\% into routes B and D. The grey line in each panel shows the CO fractional abundance, demonstrating the validity of assuming X$_{CO}$=10$^{-4}$. In the final panel, the red line shows the neutral gas temperature through the shock. }
\label{fig:models}
\end{figure*}

\begin{table}
\centering
\caption{Model parameters.}
\begin{tabular}{ll}
\hline
n(H$_2$)  	& 10$^5$ cm$^{-3}$\\
V$_s$  		& 40 km s$^{-1}$\\
t$_{sat}$   & 4.6 yr\\
T$_{n,max}$ & 4000 K \\ 
B$_0$		& 450 $\mu$G\\    
\hline
\end{tabular}
\begin{flushleft}
 Note: n($H_2$) is the pre-shock hydrogen density, V$_s$ is the shock speed, t$_{sat}$ is the saturation time, T$_{n,max}$ is the maximum temperature reached by the neutral gas.
\end{flushleft}
\label{table:models}
\end{table}
\begin{table}
\centering
\caption{Branching ratios for the freeze out routes of example sulfur bearing species in each model. Where a \# indicates a frozen species.}
\begin{tabular}{ccccc}
\hline
\textbf{Model} & \textbf{S} & \textbf{HS} & \textbf{CS}\\
\hline
Model A & 100\% \#S & 100\% \#HS    & 100\% \#CS \\\\
Model B & 100\% \#H$_2$S & 100\% \#H$_2$S    & 100\% \#HCS \\\\
Model C & 50\% \#S & 50\% \#HS    & 50\% \#CS\\
        & 50\% \#H$_2$S & 50\% \#H$_2$S & 50\%\#HCS \\ \\
Model D & 100\% \#OCS & 50\% \#HS    & 100\% \#OCS \\
        && 50\% \#H$_2$S & & \\ \\
Model E &  50\% \#OCS & 50\% \#HS    & 50\% \#OCS \\
        & 50\% \#H$_2$S & 50\% \#H$_2$S & 50\% HCS \\ 
\hline
\end{tabular}
\label{tab:networks}
\end{table}
Figure~\ref{fig:models} shows the results from four of the networks used to investigate the conditions of the shock in L1157-B1. In each case only phase II is shown, the initial abundances at log(Z)=14 are the final abundances of phase I from each network. As  explained above, sputtering occurs once t$_{sat}$ is reached, releasing mantle species into the gas phase. This sputtering is responsible for the initial large increase in abundance for each species at around log(Z)=14.8. A further increase can be seen in the H$_2$S abundance as the peak temperature is reached at log(Z)=15.8, after which H$_2$S is destroyed through gas phase reactions as the gas cools. Most of this destruction is due to the reaction, \\
\begin{center} H+H$_2$S $\rightarrow$ HS + H$_2$. \end{center} It is worth noting that the H$_2$S abundance remains low over thousands of years in cold gas if the model is allowed to continue to run after the shock passage.\par
Comparing the predicted fractional abundance of H$_2$S in each model, models B and E are the best fit. The average fractional abundance of each model is calculated as an average over the dissipation length: the whole width of the shock. The fractional abundance is given here with respect to the H$_2$ number density. Model B predicts an average H$_2$S abundance of X(H$_2$S) = 1$\times$10$^{-6}$ and model E gives X(H$_2$S) = 6$\times$10$^{-7}$. This is comparable to the observed fractional abundances of H$_2$S, as long as the H$_2$ column density used in our calculations, and inferred from CO measurements, is correct. The fractional abundance The fractional abundance of CO was taken to be X$_{CO}$=10$^{-4}$, a standard assumption which agrees with the model abundance of CO (see Fig.~\ref{fig:models}).\par
On the other hand, Models A and D predict X(H$_2$S) = 7.8$\times$10$^{-8}$ and X(H$_2$S) = 7.3$\times$10$^{-8}$ respectively. This is an order of magnitude lower than the fractional abundance calculated in Sect.~\ref{sec:abund}. \par
Furthermore, it is expected that excitation and beam effects are constant between two transitions for all emission velocities. Therefore, it can be assumed that any change in intensity ratio between the transitions of two species reflects a change in the abundances of those species. Both models A and D give varying ratios of H$_2$S to NH$_3$ which is at odds with the approximately flat intensity ratio shown in Fig.~\ref{fig:ratios}. Therefore, the best match to the data is found when at least half of the available gas phase sulfur hydrogenates as it freezes as in models B and E. \par
In contrast, CS remains at a relatively constant fractional abundance in the model throughout the shock and so one would expect a large difference between the intensities of CS and H$_2$S at high velocities, where the gas is warmer. This is consistent with the velocity limits of the detections and the ratio plots shown in Fig.~\ref{fig:ratios}. None of the H$_2$S spectra remain above the rms at more than $V=$-8 km s$^{-1}$ but \citet{gomez2015} report their CS (3-2) line extends to $V=$-19 km s$^{-1}$. We note that the differences in terminal velocities between CS and H$_2$S are not due to excitation effects but to real differences in their abundances as the CS (7-6) line, reported in the same work, reaches V=-16 km s$^{-1}$ and has E$_u$, g$_u$ and A$_{ul}$ that are similar to the H$_2$S (2$_{1,2}$-1$_{0,1}$) transition.\par
Whilst the comparison between the fractional abundances given by the models and the data is promising, it must be remembered that the networks used are limiting cases and the model is a 1D parameterization. From the results of models A and D it is clear that the observed H$_2$S emission requires that half of the sulfur on the grains be in H$_2$S. However, observations of other sulfur bearing species would have to be used to differentiate models B, C and E as the H$_2$S abundance profile appears to be largely unchanged once at least half of the frozen sulfur is hydrogentated.
\section{Summary}
\label{sec:summary}
In this work, H$_2$S in the L1157-B1 bowshock has been studied using data from the Herschel-CHESS and IRAM-30m ASAI surveys. Six detections have been reported: H$_2$S (1$_{1,0}$-1$_{0,1}$), (2$_{0,2}$-1$_{1,1}$) and (2$_{1,2}$-1$_{0,1}$); H$_2^{34}$S (1$_{1,0}$-1$_{0,1}$); and HDS (1$_{0,1}$-0$_{0,0}$) and (2$_{1,2}$-1$_{0,1}$). The main conclusions are as follows. \par
i) The H$_2$S gas in L1157-B1 has a column density of $N(H_2S)$=6.0$\pm$4.0$\times$10$^{14}$ cm$^{-2}$ and excitation temperature, T=13$\pm$6 K. This is equivalent to a fractional abundance of X(H$_2$S)$\sim$6.0$\times$10$^{-7}$. These values are based on opacity measurements using the H$_2$$^{34}$S intensity and an assumed size of 18".\par
ii) The isotopologue detections allow the deuteration fraction of H$_2$S in L1157-B1 to be calculated. A HDS:H$_2$S ratio of 2.5$\times$10$^{-2}$ is found.\par
iii) The state of sulfur on dust grains is explored by the use of a gas-grain chemical code with a C-shock where the freeze out routes of sulfur bearing species are varied in order to produce different ice compositions. These frozen species are then released into the gas by the shock and the resulting chemistry is compared to the measured abundances and intensities of molecules in L1157-B1. It is found that the best fit to the data is when at least half of each sulfur bearing species hydrogenates as it freezes. This is not in contradiction with the result of \citet{podio2014} who found OCS had to be the main sulfur bearing species on the grains to match CS and HCS$^+$ observations. This is because the 50\% hydrogenation on freeze out model remains a good fit for H$_2$S when the other 50\% of sulfur becomes OCS. Further work with a more comprehensive dataset of emission from sulfur bearing species in L1157-B1 is required to truly understand the grain composition.
\section{Acknowledgements}
J.H. is funded by an STFC studentship. L.P. has received funding from the European Union Seventh Framework Programme (FP7/2007-2013) under grant agreement No. 267251. I.J.-S. acknowledges the financial support received from the STFC through an Ernest Rutherford Fellowship (proposal number ST/L004801/1).



\bibliographystyle{mnras}
\bibliography{h2s}







\bsp	
\label{lastpage}
\end{document}